\documentclass[letterpaper, 10 pt, conference]{ieeeconf}  % Comment this line out if you need a4paper

\usepackage{amsmath,amsfonts}
\usepackage{algorithmic}
\usepackage{algorithm}
\usepackage{multirow}
\usepackage[table,xcdraw]{xcolor}
\usepackage{amsfonts}
\usepackage{array}
\usepackage{comment}
\usepackage{gensymb}
\usepackage{textcomp}
\usepackage{stfloats}
\usepackage{url}
\usepackage{verbatim}
\usepackage{graphicx}
\usepackage{cite}

\IEEEoverridecommandlockouts                              % This command is only needed if 
\title{\LARGE \bf
Brain Connectivity Features-based Age Group Classification using Temporal Asynchrony Audio-Visual Integration Task
}

\author{Prerna Singh$^1$, Ayush Tripathi$^2$,  Lalan Kumar$^3$, and Tapan Kumar Gandhi$^4$
\thanks{$^{1}$Prerna Singh is with the Bharti School of Telecommunication Technology and Management, Indian Institute of Technology Delhi, New Delhi - 110016, India. {\tt\small bsz208534@dbst.iitd.ac.in}}%
\thanks{$^{2}$Ayush Tripathi is with the Department of Electrical Engineering, Indian Institute of Technology Delhi, New Delhi - 110016, India and is supported by the Prime Minister Research Fellows (PMRF) Scheme with grant number PLN08/PMRF. {\tt\small eez208477@ee.iitd.ac.in}}%
\thanks{$^{3}$Lalan Kumar is with the Department of Electrical Engineering, Bharti School of Telecommunication Technology and Management, and Yardi School of Artificial Intelligence, Indian Institute of Technology Delhi, New Delhi - 110016, India. {\tt\small lalank@ee.iitd.ac.in}}%
\thanks{$^{4}$Tapan Kumar Gandhi is with the Department of Electrical Engineering, and Bharti School of Telecommunication Technology and Management, Indian Institute of Technology Delhi, New Delhi - 110016, India. {\tt\small tgandhi@ee.iitd.ac.in}}%
}

\begin{document}

\maketitle
\thispagestyle{empty}
\pagestyle{empty}

%%%%%%%%%%%%%%%%%%%%%%%%%%%%%%%%%%%%%%%%%%%%%%%%%%%%%%%%%%%%%%%%%%%%%%%%%%%%%%%%
\begin{abstract}

The process of integration of inputs from several sensory modalities in the human brain is referred to as multisensory integration. Age-related cognitive decline leads to a loss in the ability of the brain to conceive multisensory inputs. There has been considerable work done in the study of such cognitive changes for the old age groups. However, in the case of middle age groups, such analysis is limited. Motivated by this, in the current work, EEG-based functional connectivity during audiovisual temporal asynchrony integration task for middle-aged groups is explored. Investigation has been carried out during different tasks such as: unimodal audio, unimodal visual, and variations of audio-visual stimulus. A correlation-based functional connectivity analysis is done, and the changes among different age groups including: young (18-25 years), transition from young to medium age (25-33 years), and medium (33-41 years), are observed. Furthermore, features extracted from the connectivity graphs have been used to classify among the different age groups. Classification accuracies of $89.4\%$ and $88.4\%$ are obtained for the Audio and Audio-50-Visual stimuli cases with a Random Forest based classifier, thereby validating the efficacy of the proposed method.

%Age-related functional connectivity in middle-aged groups during audiovisual temporal asynchrony integration tasks remains unexplored. Electroencephalogram (EEG) of 5 young participants, Y (18-25 years), 5 participants transitioning from young to medium age group, T (25-33 years), and 5 medium aged participants, M (33-41 years) were recorded during an audiovisual temporal asynchrony integration task with five conditions [auditory (A), visual (V), AV, A50V, and V50A]. For these age groups, correlation-based functional connectivity was investigated using NetworkX. In the case of A and A50V stimuli, functional connectivity changes drastically in midlife, primarily in the central, pre-frontal, parietal, and superior temporal regions of the brain. Using the clustering coefficient and centrality features, the three age groups can be classified. RF has given a maximum accuracy of 89.4 \% in the case of 'A' stimuli at a correlation threshold value of 0.8. which can be compensated by A50V stimuli in the elderly.
\end{abstract}

%%%%%%%%%%%%%%%%%%%%%%%%%%%%%%%%%%%%%%%%%%%%%%%%%%%%%%%%%%%%%%%%%%%%%%%%%%%%%%%%
\section{Introduction}
Daily behavior of humans is shaped by multisensory integration which facilitates and enhances perceptual abilities. Multisensory integration is the process of integrating stimuli from several sensory channels such as audio and visual into common perceptual states \cite{ernst2004merging}. The human sensory systems are tremendously affected with age thereby leading to loss of functionality with growing age. This in turn affects the ability to conceive multisensory inputs. Such cognitive losses brought on by aging can, however, be successfully compensated by the integration of multisensory input \cite{lovelace2003irrelevant,molholm2004multisensory}. The most prevalent multisensory integration modality in both humans and animals is Audiovisual Integration (AVI). AVI happens considerably only when there is less than 100 ms between the appearance of auditory and visual stimuli \cite{stein2012new,ren2018comparison,ren2017audiovisual}. Based on the temporal nature of AVI, synchronous and asynchronous AVI tasks can be studied with respect to age.

AVI occurs mainly in the superior temporal regions and the occipital regions \cite{falchier2002anatomical}. A major concern in audiovisual integration is to analyze the effect of integration and coordination of widely scattered brain areas on the perceptual functions. In this regard, the analysis of functional connectivity is an appealing prospect. Such an analysis involves identifying temporal correlations or synchronization in physiological signals collected from distinct brain areas \cite{fingelkurts2005functional,stam2009graph,wang2018mulan}. In literature, Graph theoretical analysis has been used to examine functional connectivity in EEG data \cite{vecchio2014human}. Several studies have employed graph theoretical network topography analysis to study age-related audio-visual integration in EEG data \cite{wang2017beta}. However, most of the prior work has been focused on examining the functional connectivity in old age groups. As a result, the changes in functional connectivity and network efficiency during audiovisual integration tasks in the middle or medium age group is widely unexplored. Detecting disorders such as mild cognitive impairment (MCI) in middle age may aid in the early detection of Alzheimer's disease (AD). Even though the pathophysiological process begins 10-15 years before the development of dementia, investigations on middle-age groups are sparse \cite{kremen2014early}. Shifting the focus to MCI identification in middle-aged group would increase the possibility of identifying early biomarkers of cognitive decline. 

Motivated by this, the current study focuses on examining the brain network connectivity of healthy participants in their middle ages. To investigate changes in functional connectivity during audiovisual integration tasks in middle-aged adults (in synchrony and asynchrony),  an auditory or visual stimuli discrimination task was designed. The experiment comprised of three types of stimuli: unimodal visual (V), unimodal auditory (A), and bimodal audio-visual (AV) stimuli. The work focuses on three subject groups: young (Y) 18-25 years, the transition from young to medium age (T), 25-33 years, and medium age (M), 33-41 years. EEG signals are captured from distinct brain areas and the corresponding functional connectivity network is analysed. Furthermore, several features from these networks are extracted. These are used in conjunction with different Machine Learning models for classifying into different age groups.   %We next utilized graph analytics parameters like centrality and clustering coefficient to investigate how functional connectivity varies for these age groups for different audio-visual stimuli. Later we classified these three age groups using different classifiers based on graph analytics features. 

\section{Materials and Methods}

\subsection{Participants}
In this study, Fifteen healthy participants from three distinct age groups : young subjects (18-24 years, with mean age ± SD, 21 ± 1.41 years, n=5), transition from young to medium age subjects (25-32 years, with mean age ± SD, 27.4 ± 2.57 years, n=5), and medium age subjects (33-41 years, with mean age ± SD, 37 ± 2.8 years, n=5) were recruited with their consent. All the participants were unaware of the experiment's goals and had normal hearing with normal or corrected to normal vision. All the participants had normal Mini-Mental State Examination (MMSE) score ($>$24) and had no history of cognitive impairment \cite{bravo1997age}. The data collection protocol was approved by the Institute Ethics Committee of IIT Delhi (Reference $\#2021/P052$).

\begin{figure}[!t]
    \centering{
    \vspace{5pt}
    \includegraphics[width=0.7\linewidth]{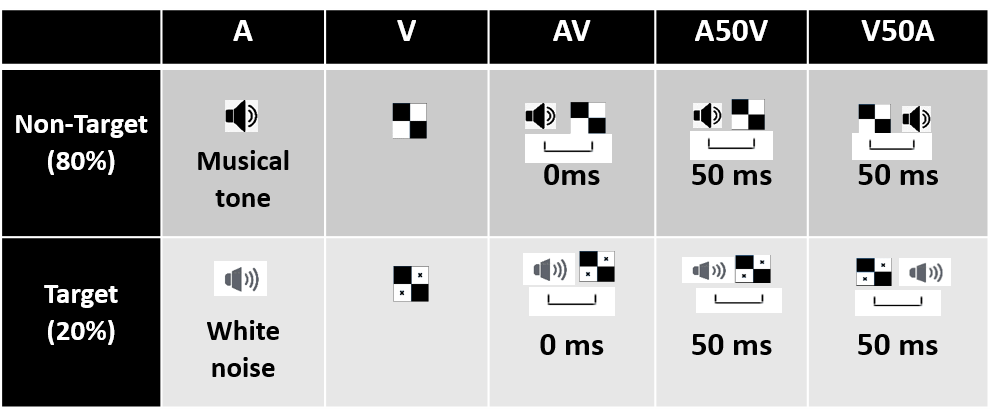}
    \caption{Description of different stimulus involved during data collection.}
    \label{fig:Stimuli description}
    }
\end{figure}

\begin{figure}[!t]
    \centering{
    \includegraphics[width=0.75\linewidth]{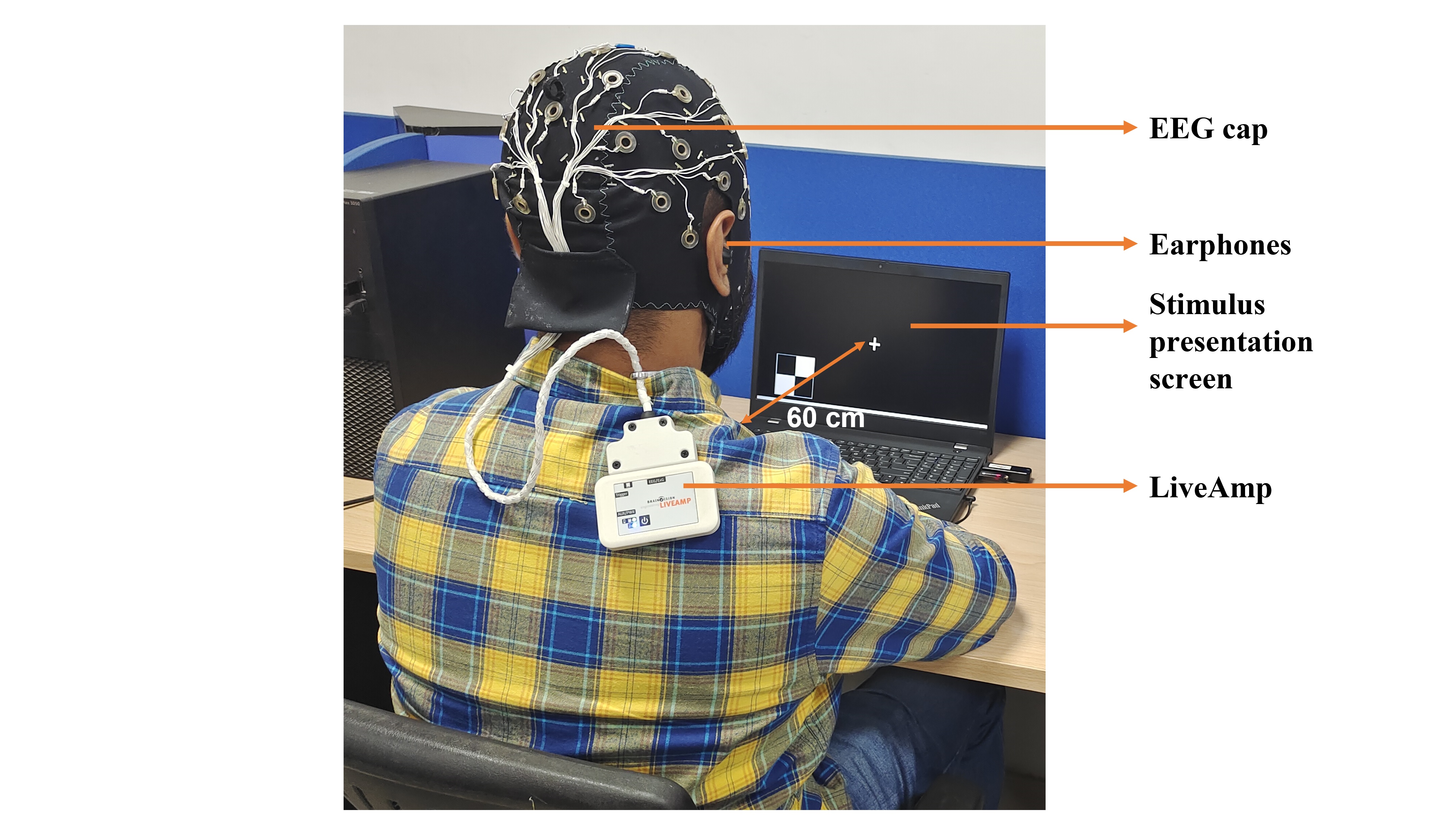}
    \caption{Depiction of the experimental setup.}
    \label{fig:Experimental setup}
    }
\end{figure}

\subsection{Stimuli and Task}
For the experiments, two different kinds of auditory and visual stimulus, namely: target and non-target were used. The non-target target visual stimulus was a black and white checkered box, while the target visual stimulus contained two cross markings within the checkered box (52 mm × 52 mm, with a visual angle of 5\degree). The non-target and target auditory stimulus were a musical sound, and white noise respectively. For 150 ms, visual stimuli (V) was randomly delivered to the lower left or lower right quadrant of a black screen on a 15.6-inch laptop screen situated 60 cm in front of the participant's eyes. Auditory stimuli (A) was delivered to either the left or right ear through earphones at a sound pressure level of 60 dB for 150 ms.

The experiment comprised of unimodal audio (A), unimodal visual (V), and audio-visual stimulus. For the audio-visual case, stimulus was presented in three ways depending on the duration of stimulus onset asynchrony (SOA). These three cases are: simultaneous audio-visual (AV), auditory lag visual by 50 ms (V50A), and auditory lead visual by 50 ms (A50V). The length of each trial of each stimulus ranged between 150 and 250 ms, depending on the SOA determined based on prior behavioural investigations \cite{yang2014elevated}. Each participant had five sessions, which began with a 3000 ms fixation time. Subsequently, 20 stimulus of each of the five types (A, V, AV, A50V, V50A) appeared on the screen. For each condition on the left or right side, there were 80\% non-target stimuli and 20\% target stimuli. All stimuli were delivered with an inter stimulus interval ranging from 1300 to 1800 ms. Participants were asked to find out whether the targets appeared on the left or right side of the screen as soon and precisely as possible. A description of the different kinds of stimulus is provided in Figure \ref{fig:Stimuli description} and a representative data collection setup is presented in Figure \ref{fig:Experimental setup}.

%circumstances, each with a distinct duration of stimulus onset asynchrony (SOA) of 0 ms and 50 ms: simultaneous audio-visual (AV); auditory lag visual 50 ms (V50A); or auditory leading visual 50 ms (A50V). 

\subsection{EEG Data Collection and Pre-Processing}

EEG data was recorded using 31 scalp electrodes mounted on an electrode cap (EasyCap from Brain Products) and placed in accordance with the International 10-20 electrode placement standards. EEG signals were amplified and digitized at a sampling rate of 500 Hz using LiveAmp (Brain Products) amplifier. The impedance of each EEG electrode was maintained below $20 k\Omega$. 
 
The recorded EEG data was pre-processed and analyzed using EEGLAB toolbox \cite{delorme2004eeglab} in MATLAB R2021a. First, the signals were filtered between 0.5 and 40 Hz and re-referenced to average reference. Blinks, eye movements, power line interference (50 Hz), electrocardiographic artifacts, and muscular artifacts were visually detected and removed using independent component analysis (ICA) \cite{delorme2004eeglab}. Subsequently, data was separated into epochs with time point ranging from 500 ms before stimulus onset to 900 ms after stimulus start, and baseline correction was applied. This resulted in a total of $100$ epochs (including both target and non-target stimulus) comprising of $700$ time points of each stimulus type per participant.

\section{Experimental Details}

\subsection{Graph Construction}

\begin{figure}[!t]
    \centering{
    \vspace{5pt}
    \includegraphics[width=0.85\linewidth]{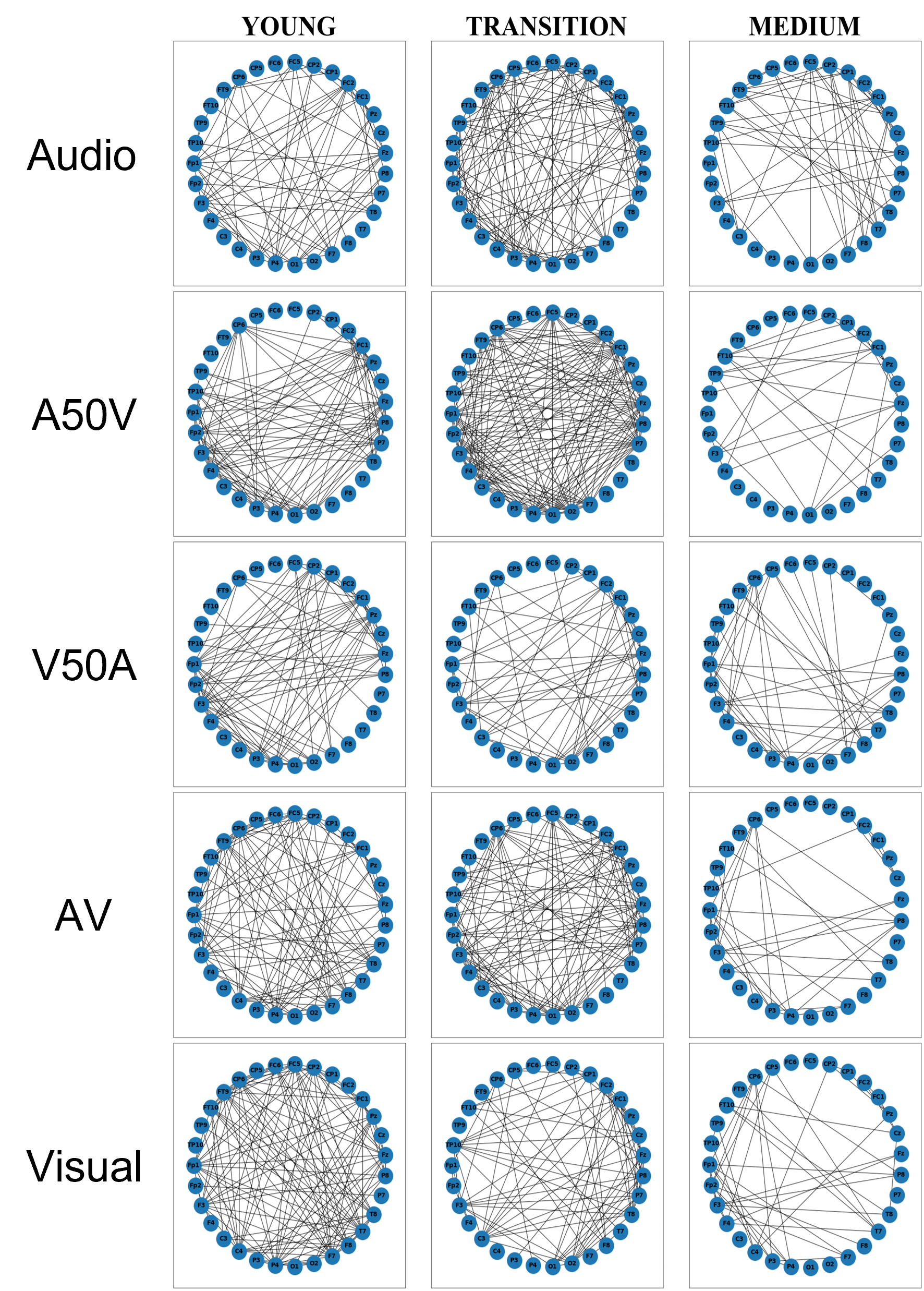}
    \caption{Graphs constructed by using averaged epochs for different age-groups and stimulus types with correlation threshold $\rho_{th}=0.8$}
    \label{fig:avergaed_graph}
    }
\end{figure}

The preprocessed EEG signals after epoch formation are used to construct corresponding functional connectivity graph. Each epoch can be represented as a $700 \times 31$ dimensional matrix ($V$), where $700$ denotes the number of time steps and $31$ represents the number of channels. First, a $31 \times 31$ dimensional adjacency matrix ($\hat{A}$) is constructed based on Pearson correlation between signals at different pairs of EEG channels. Let $\mu_{V(:,i)}$ denote the mean of the $i^{th}$ EEG channel, then, the elements of the adjacency matrix $\hat{A}$ are computed as,

\begin{equation}
    \hat{A}(i,j) = \frac{\sum_{l=1}^{700} (V(l,i)-\mu_{V(:,i)})(V(l,j)-\mu_{V(:,j)})}{\sqrt{\sum_{l=1}^{700} (V(l,i)-\mu_{V(:,i)})^2 \sum_{l=1}^{700}(V(l,j)-\mu_{V(:,j)})^2}}
\end{equation}

Subsequently, a binary adjacency matrix ($A$) is constructed by using the elements of $\hat{A}$ as follows,

\begin{equation}
A(i,j) =  
    \begin{cases}
    1 ,& \text{if } i\neq j \,\, \text{and } \hat{A}(i,j)\geq \rho_{th}\\
    0 ,& \text{otherwise } 
\end{cases}
\end{equation}

The resulting adjacency matrix forms the connectivity graph for a given epoch. The threshold parameter ($\rho_{th}$), controls the strength of correlation required for the formation of an edge between two nodes. A low value of $\rho_{th}$ would result in the large number of edges, while a high value would result in obtaining edges between only those nodes which have a sufficiently high correlation between their signals. Sample graphs constructed by using average epochs of all the subjects for different age groups and different types of stimulus for a threshold of $0.8$ are presented in Figure \ref{fig:avergaed_graph}.

\subsection{Feature Extraction}

From the computed adjacency matrix, $5$ different node-level features are extracted for the task of age-group classification. This results in a total of $5 \times 31 = 155$ features for the connectivity graph of a given epoch. The following features are extracted using the \textit{NetworkX} package \cite{SciPyProceedings_11}: 

\begin{enumerate}
    \item Degree Centrality: For a node $v$, degree centrality is a measure of the number of nodes connected to it. Mathematically, it is computed as,
    \begin{equation}
        d_v = \frac{\sum_{l=1}^{N}A(v,l)}{N-1}
    \end{equation}
    where, $N$ is the total number of nodes in the graph.

    \item Betweenness Centrality: It is a measure of importance of a node in the graph by measuring the number of times a given node $v$ occurs in shortest path between all pairs of nodes. It is computed as,
    \begin{equation}
        b_v = \sum_{a,b\in V}\frac{\sigma(a,b|v)}{\sigma(a,b)}
    \end{equation}
    where, $V$ is the set of nodes, $\sigma(a,b)$ is the number of shortest paths between nodes $a$ and $b$, and $\sigma(a,b|v)$ are such shortest paths passing through node $v$. %If $a=b$ then $\sigma(a,b)=1$, and if $v\in a,b$ then $\sigma(a,b|v) = 0$.

    \item Eigenvector Centrality: It is a measure of centrality of a node $v$ based on the centrality of neighboring nodes. It is computed as the $v^{th}$ element of the vector $e$ computed from the equation:
    \begin{equation}
        Ae = \lambda e
    \end{equation}
    Since, all the entries of matrix $A$ are non-negative valued, there exists a unique solution $e$, all the entries of which are positive for the largest eigenvector $\lambda$.

    \item Closeness Centrality: For a node $v$, it is defined as the reciprocal of average shortest distance of the node over all reachable nodes. Mathematically, it is computed as,
    \begin{equation}
        c_v = \frac{n-1}{\sum_{l=1}^{n-1}d(v,l)}
    \end{equation}
    where, $d(v, l)$ is the shortest-path distance between nodes $v$ and $l$, and $n$ is the number of nodes reachable from node $v$.
    
    \item Clustering Coefficient: The clustering coefficient of a node $v$ is defined as the fraction of possible triangles through that node. Mathematically, it is computed as:
    \begin{equation}
    \kappa_v = \frac{T(v)}{(deg(v))(deg(v)-1)}
    \end{equation}
    where, $T(v)$ is the number of triangles through node $v$ and $deg(v)$ is the number of edges connected to it.

\end{enumerate}

\subsection{Age-group classification}

Based on the feature extraction procedure, a total of $155$ features per epoch are obtained. The recorded dataset comprises of $1500$ such epochs per stimulus type equally divided among three classes (Y, T, and M). In order to validate the efficacy of the extracted features in classifying the data among different age groups, a $10$-fold cross validation strategy is adopted and mean accuracy scores are reported. In the first set of experiments, the effect of variation of the correlation threshold parameter ($\rho_{th}$) on the classification performance by using Random Forest (RF) classifier for different stimulus types is analyzed. Subsequently, the performance of different classifiers, namely: Support Vector Machines with linear kernel (Linear SVM), Logistic Regression (LR) and k-Nearest Neighbors (kNN) for different types of stimulus with $\rho_{th}=0.8$ is also analyzed. For all the classifiers, default parameters from the scikit-learn\cite{sklearn_api2} library are used.

\section{Results}

\subsection{Brain connectivity}
Using a $\rho_{th}$ value of 0.8, the generated brain functional connectivity plots for different age groups are illustrated in Figure \ref{fig:avergaed_graph}. The nodes in the graph represent different EEG electrodes and edges are used to represent connection between the electrode locations to form the functional connectivity graph. As depicted in Figure \ref{fig:avergaed_graph}, there is a change in the number of edges for different age groups. Among all stimuli, A50V and A had the largest reduction in connectivity from the T to M phase in the Central, Prefrontal, Occipital, and Superior temporal areas which play a central role in AVI in the brain. In case of V and V50A, the Y age group has the highest edge count, followed by the T and M age groups. This indicates that features extracted from the nodes can be used as a marker for age-group classification. 

\begin{figure}[!t]
    \centering{
    \vspace{5pt}
    \includegraphics[width=0.85\linewidth]{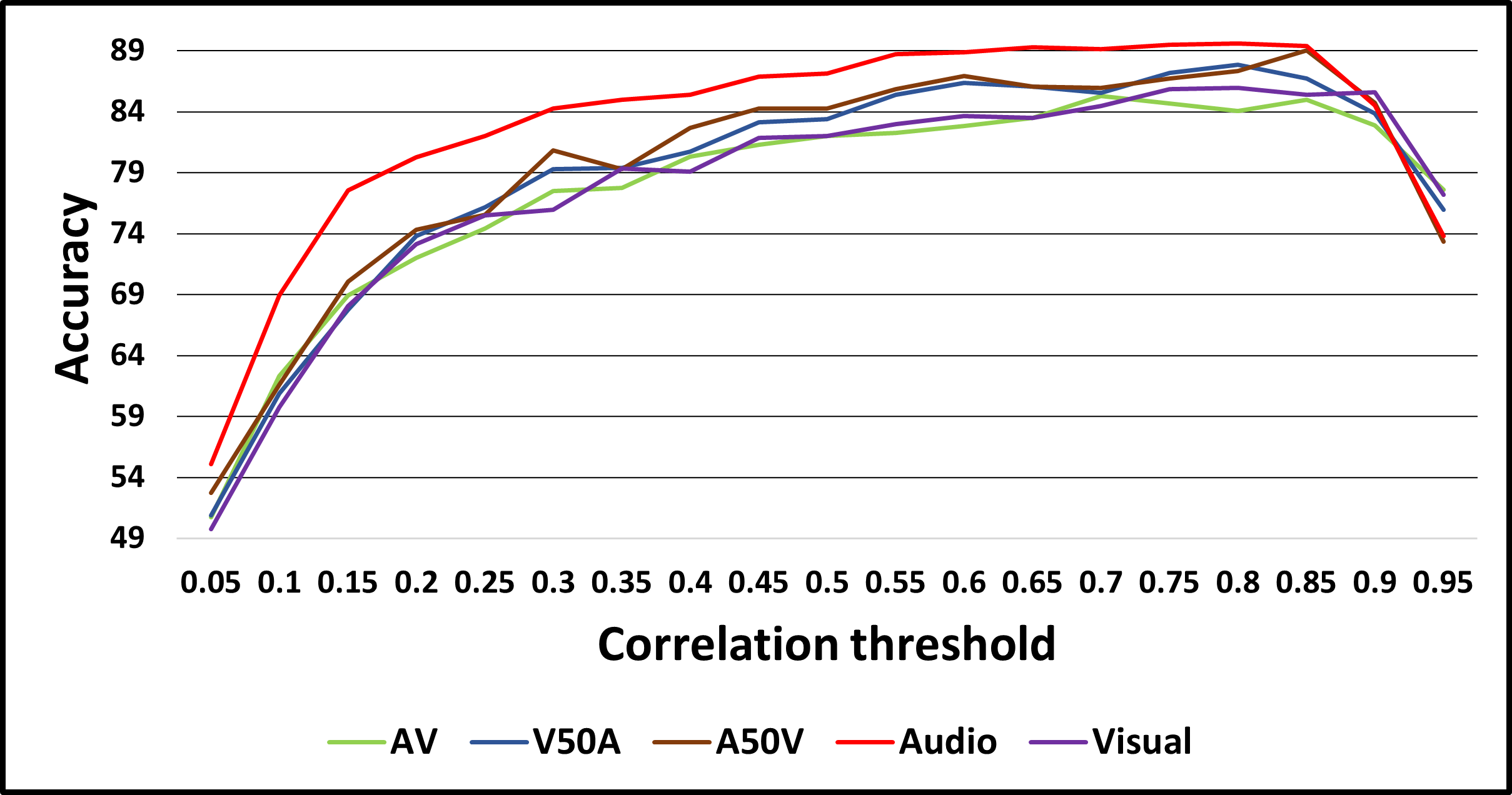}
    \caption{The effect of correlation threshold parameter ($\rho_{th}$) on recognition accuracies for different types of stimulus using RF classifier}
    \label{fig:threshold_variation}
    }
\end{figure}

\begin{table}[!t]
\caption{Recognition accuracies by using different classification algorithms for different stimulus types.}
\centering
\scalebox{0.8}{
\begin{tabular}{ccccc}
\hline 
\textbf{Stimulus} & \textbf{LR} & \textbf{LinearSVM} & \textbf{kNN} & \textbf{RF}                           \\\hline\hline
\textbf{AV}       & 76.00       & 76.93              & 74.53        & {\color[HTML]{0000FF} 84.20}          \\
\textbf{V50A}     & 80.73       & 82.00              & 80.00        & {\color[HTML]{0000FF} 87.47}          \\
\textbf{A50V}     & 79.73       & 81.47              & 78.93        & {\color[HTML]{0000FF} 88.40}          \\
\textbf{Audio}    & 80.87       & 82.80              & 81.60        & {\color[HTML]{0000FF} \textbf{89.40}} \\
\textbf{Visual}   & 78.27       & 79.87              & 76.67        & {\color[HTML]{0000FF} 86.53} \\\hline        
\end{tabular}
}
\label{tab:results}
\end{table}

\subsection{Classification accuracy}

The effect of correlation threshold ($\rho_{th}$) on the classification accuracies for different types of stimulus is plotted in Figure \ref{fig:threshold_variation}. It may be noted that the accuracy first shows an increasing trend before dipping off. The optimal value of $\rho_{th}$ is observed to be $0.8$ and this value is retained subsequently. With a fixed correlation threshold, $\rho_{th}=0.8$, different classifiers were used for the age-group classification task and the results are presented in Table \ref{tab:results}. Random Forest based classifier yielded the the highest classification accuracy of 89.4\% in the Audio stimuli case, followed by 88.4\% for A50V stimuli. This result is in accordance with the temporal principle of audio-visual integration. It shows that hearing loss in old age can be compensated by the integration of audio and visual stimuli with a time lag of 50 milliseconds.

\section{Conclusion}

In this study, the changes in brain connectivity network with different age-group is analysed. It is found that Audiovisual Integration (AVI) induced by peripherally delivered AV stimuli varies with age. This study also reveals that correlation-based functional connectivity in the EEG signals owing to peripherally delivered stimuli declines in midlife. AVI causes changes in brain connectivity that begin to reflect while transitioning from the young (Y) to the Medium (M) age group. It is found that by using functional connectivity-based features for different AVI stimuli, three age groups: young (Y), transition from young to medium (T), and Medium (M), can be categorized. Future work may be focused on exploring discriminative features that can be utilized for the task of EEG-based age group classification including the old age population.

\bibliographystyle{IEEEtran}
\bibliography{refs.bib}

\end{document}